\documentclass[sigconf]{acmart}

\setcopyright{none}
\acmConference[Technical Report]{}{December}{2019}
\usepackage{xspace}
\usepackage{textcomp}
\usepackage{soul}
\usepackage{marginnote}
\usepackage{hyperref}
\usepackage{amsfonts}
\usepackage{graphicx}
\usepackage{float}
\usepackage{subfig,rotate}
\usepackage{verbatim}
\usepackage{comment}
\usepackage{versions}
\usepackage{multirow}
\usepackage{longtable}
\usepackage{booktabs}
\usepackage{makecell}
\usepackage{caption}
\usepackage{acro}
\usepackage{tabularx}
\usepackage{adjustbox}
\usepackage{listings}
\usepackage{color}
\usepackage{stmaryrd}
\usepackage{amsmath}
\usepackage{array}
\usepackage[boxed]{algorithm2e}

\def\bitcoin{%
  \leavevmode
  \vtop{\offinterlineskip 
    \setbox0=\hbox{B}%
    \setbox2=\hbox to\wd0{\hfil\hskip-.03em
    \vrule height .3ex width .15ex\hskip .08em
    \vrule height .3ex width .15ex\hfil}
    \vbox{\copy2\box0}\box2}}

\definecolor{dkgreen}{rgb}{0,0.6,0}
\definecolor{gray}{rgb}{0.5,0.5,0.5}
\definecolor{mauve}{rgb}{0.58,0,0.82}

\DeclareCaptionFormat{listing}{#1#2#3}
\begin{document}

\title[Investigating MMM Ponzi scheme on Bitcoin]{Investigating MMM Ponzi scheme on Bitcoin}

\author{Yazan Boshmaf}
\affiliation{
  \institution{Qatar Computing Research Institute, HBKU}
}
\author{Charitha Elvitigala}
\affiliation{
  \institution{University of Colombo}
}
\author{Husam Al Jawaheri}
\affiliation{
  \institution{University of Luxembourg}
}
\author{Primal Wijesekera}
\affiliation{
  \institution{University of California, Berkeley}
}
\author{Mashael Al Sabah}
\affiliation{
  \institution{Qatar Computing Research Institute, HBKU}
}
\renewcommand{\shortauthors}{Boshmaf et al.}

\begin{abstract}
Cybercriminals exploit cryptocurrencies to carry out illicit activities. In this paper, we focus on Ponzi schemes that operate on Bitcoin and perform an in-depth analysis of MMM, one of the oldest and most popular Ponzi schemes. Based on 423K transactions involving 16K addresses, we show that: (1) Starting Sep 2014, the scheme goes through three phases over three years. At its peak, MMM circulated more than 150M dollars a day, after which it collapsed by the end of Jun 2016. (2) There is a high income inequality between MMM members, with the daily Gini index reaching more than 0.9. The scheme also exhibits a zero-sum investment model, in which one member's loss is another member's gain. The percentage of victims who never made any profit has grown from 0\% to 41\% in five months, during which the top-earning scammer has made 765K dollars in profit. (3) The scheme has a global reach with 80 different member countries but a highly-asymmetrical flow of money between them. While India and Indonesia have the largest pairwise flow in MMM, members in Indonesia have received 12x more money than they have sent to their counterparts in India.
\end{abstract}

\maketitle

\section{Introduction}
\label{sec:introduction}

Cryptocurrencies represent an alternative to the centralized, highly-regulated financial systems~\cite{nakamoto2008bitcoin}. Bitcoin, for example, facilitates sending and receiving money without relying on third-parties, such as banks. However, the decentralized control that underpins cryptocurrencies has garnered unwanted attention from cybercriminals, who exploit them to evade financial regulations and engage in various illicit activities, including financial fraud~\cite{tasca2015digital,dion2013ll,vasek2015there} and money laundering~\cite{money_laundering,reuter2005chasing,bitcoin_laundry_services}.

In this paper, we focus on Ponzi schemes that use Bitcoin for their operations~\cite{vasek2018analyzing}. In concept, a Ponzi scheme, also known as a high-yield investment program (HYIP), is a fraudulent investing scam promising high rates of return with low risk to investors. The return on investment (ROI) for early investors in a Ponzi scheme is generated by acquiring new investors. While online Ponzi schemes have been studied for a while~\cite{artzrouni2009mathematics,moore2012postmodern,cross2018anatomy}, the use of Bitcoin introduces unique challenges that make them harder to investigation~\cite{bartoletti2017dissecting, bartoletti2018data, moore2012postmodern}. In particular, it is unclear how to identify and characterize the Ponzi operation, given Bitcoin's pseudo-anonymous privacy model~\cite{nakamoto2008bitcoin}. We bridge this knowledge gap and present a method to investigate the operation of Ponzi schemes by linking and analyzing data collected from multiple public sources. We apply this method in the real-world and characterize the operation of MMM~\cite{cross2018anatomy}, one of the oldest and most popular Ponzi schemes on Bitcoin (\S\ref{sec:background}).

We crawled 2.5M user profiles on BitcoinTalk,\footnote{\url{https://bitcointalk.org}} a famous discussion forum, searching for Bitcoin addresses associated with MMM members, ending up with 15,736 addresses. We then extracted the corresponding transactions from Bitcoin's blockchain, ending up with 422,953 transaction (\S\ref{sec:dataset}).

Using the collected dataset, we analyzed the lifecycle of MMM in terms of its daily transaction volume and money flow. In particular, we studied the scheme under three formally-defined phases based on volume, namely bootstrap, hyperoperation, and collapse, under which MMM operated for three years starting on Sep 1, 2014. At its peak, the scheme circulated more than 150M dollars a day, before starting to collapse on Jun 29, 2016. We also analyzed the fraudulent operation of MMM, focusing on income inequality among members, return on investment, and member classification. We computed the daily Gini index~\cite{gastwirth1972estimation}, which measures the distribution of income across a group of people, and found a highly-skewed concentration, with values reaches more than 0.9 and an overall average of 0.65. The scheme also exhibited a Ponzi zero-sum investment model, in which a member's profit is another member's loss, except for nearly 73M dollars that were sent to unknown addresses. The percentage of victims, who are members that never made a profit, increased from 0\% to 41\% in five months, after which the scheme started to collapse, with the top-earning scammer making 765K dollars in profit. We then analyzed the external flow of money in which money is sent to or receive from MMM addresses as deposit or withdrawal transactions, respectively. We found that most of the deposits (85.54\%) and withdrawals (91.57\%) are associated with unknown addresses, but for those which we were able to identify, the highest percentages were associated with exchange services, with 4.8x more withdrawals than deposits. Finally, we analyzed the geopolitics of MMM, focusing on how much money flows between countries. Among 80 different member countries, we found that the largest flow of money, which was between India and Indonesia, is 10x larger than the smallest flow, which was between Thailand and Taiwan. This pairwise money flow, however, is highly asymmetrical and could impact local economies. For example, members in Indonesia have received 12x more money than they have sent to their counterparts in India (\S\ref{sec:analysis}).

To this end, we discuss the implications of our findings for cryptocurrency regulations, economy, and user privacy (\S\ref{sec:discussion}). We summarize our contributions in the following:
\begin{enumerate}
\item We propose an analytical method and a set of metrics which are useful in investigating Ponzi schemes that operate with cryptocurrencies like Bitcoin. 
\item We present the first in-depth, temporal characterization of a real-world Ponzi scheme on Bitcoin.
\item We share a unique dataset of Bitcoin transactions and addresses which are associated with MMM.
\end{enumerate}

\section{Background}
\label{sec:background}

We now present the background for this work.

\subsection{Bitcoin}
\label{sec:background_bitcoin}

Bitcoin is the first and the most popular cryptocurrency network~\cite{nakamoto2008bitcoin}. In Bitcoin, the identity of a user is hidden by using public pseudonyms called addresses. A Bitcoin address is an alphanumeric identifier that is derived from the public key of a public/private key pair. It has three different formats that are in use today, each defining an address type, as shown in Table~\ref{tab:bitcoin_regex}.

\begin{table}
\caption{Bitcoin address types and regex patterns.}
\centering
{\small
\begin{tabular}{llr}\toprule
    Type & Description & Regex pattern\\ \midrule
    P2PKH & Pay to public key hash & \texttt{1[a-km-zA-HJ-NP-Z1-9]\{25,34\}}\\
    P2SH& Pay to script hash & \texttt{3[a-km-zA-HJ-NP-Z1-9]\{25,34\}}\\
    Bech32 & Segregated witness & \texttt{bc1[a-zA-HJ-NP-Z0-9]\{25,39\}}\\
\bottomrule
\end{tabular}
}
\label{tab:bitcoin_regex}   
\end{table}

The set of public/private keys that are owned by a user is called a wallet. Private keys are used to sign inputs of transactions as a proof of ownership. To protect a transaction, the user sending the money signs the transaction using her private keys, and then broadcasts it to the network for verification.

Bitcoin transactions are stored in a public, decentralized ledger called a blockchain, which means anyone could try to identify user transactions by analyzing the blockchain. Doing so, however, is difficult since user identities are not recorded in the blockchain, only Bitcoin addresses. As such, Bitcoin's privacy model is pseudo-anonymous; if user identities are linked to Bitcoin addresses, their transactions can be identified. This is the case when users publicly post their Bitcoin addresses along with their personally identifiable information (PII), commonly found in forums like BitcoinTalk.

It is possible to exchange bitcoins with fiat currencies through centralized cryptocurrency exchange services, such as Luno\footnote{\url{https://www.luno.com}} and Huobi,\footnote{\url{https://www.huobi.com}} or decentralized ones, such as WavesDex.\footnote{\url{https://client.wavesplatform.com}} While largely unregulated, the price is often based on supply and demand, but can still vary significantly during a trading day. Services like CoinDesk\footnote{\url{https://www.coindesk.com}} aggregate pricing data from exchanges and provide market statistics, such as open/close and high/low values for a trading day.

\subsection{MMM}
\label{sec:background_mmm}

MMM is a Russian Ponzi scheme which was founded in the 1990s~\cite{cross2018anatomy}. It operated as an investment fund and has defrauded millions of people around the globe by promising dividends of up to 3,000\% a year. In Jul 1994, the Russian police shut down MMM for illegally issuing unregistered securities. The scheme, however, reopened in 2011 as MMM Global,\footnote{\url{https://mmmglobal.world}} with subsidiaries in 110 countries. MMM is popular in Africa~\cite{esoimeme2018money} and Asia~\cite{purba2017economics}, which is partly attributed to poverty, lack of regulations or law enforcement, and limited access to financial institution such as banks.

\begin{figure}[t]
	\centering
	\includegraphics[width=0.95\linewidth]{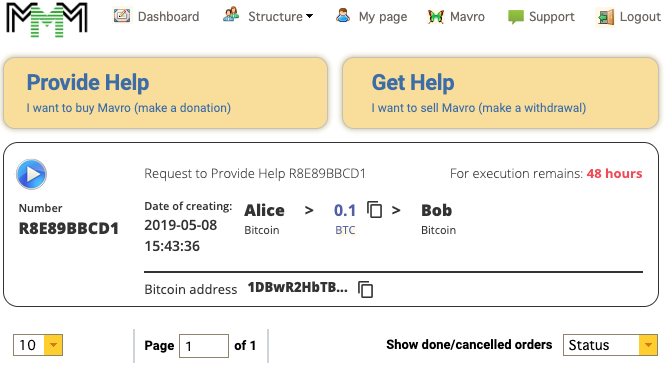}
	\caption{MMM member dashboard showing a help request.}
	\label{fig:mmm_dashboard}
\end{figure}

MMM operates as a mutual aid fund where people help each other. A user can get financial help of up to 10K dollars worth of bitcoins. Providing and getting help, however, is done in MMM's virtual currency called Mavro, which is pegged with Bitcoin at a fixed exchange rate of 1 bitcoin = 1 mavro. New MMM members start with no mavro, and the only way to buy it is to provide help, as shown in Figure~\ref{fig:mmm_dashboard}. Once help is provided in the form of a Bitcoin transaction, the user receives mavro that is equivalent to the input value of the transaction, in addition to a 30\% monthly return. For example, when Alice helps Bob by transferring 0.1 bitcoins, she immediately receives 0.1 mavro, which will turn into 0.13 mavro in a month if she does not sell them. Members can opt into a 100\% return rate by joining MMM Extra and perform simple daily tasks to promote MMM's global operations, such as liking YouTube videos, posting on Facebook groups, and joining BitcoinTalk forums.

\section{Dataset}
\label{sec:dataset}

We crawled BitcoinTalk to search for users whose profiles contain Bitcoin addresses and also point to their member profiles on MMM, ending up with 15,736 unique user$\shortrightarrow$address$\shortrightarrow$member matches out of 2.5M user profiles.\footnote{For privacy reasons, we replace user identifiers with pseudonyms, such as Alice and Bob, and show only the first 10 characters of any of their Bitcoin addresses or wallets.} We then used BlockSci~\cite{kalodner2017blocksci} to collect the transactions which involve addresses associated with MMM members from Bitcoin's blockchain, ending up with 422,953 transactions.

Our goal is to collect Bitcoin addresses and transactions which are highly-likely to be associated with MMM members. As such, we follow stringent validation rules that consequently lead to a smaller dataset and lower estimates of how much money is involved. This dataset size vs. quality tradeoff is typical in e-crime investigations and forensics~\cite{garfinkel2010digital}, where a small but reliable sample is considered better than a large but noisy one, as long as the underlaying properties are preserved and captured. While we focus on collecting a high-quality dataset, we discuss ways to increase its size in~\S\ref{sec:discussion}.

\subsection{Ponzi addresses}
\label{sec:dataset_addresses}

BitcoinTalk is one of the most popular Bitcoin discussion forums. As of May 2019, it hosted more than 50M posts. Given its popularity, we used it as a source to collect Bitcoin addresses that are associated with MMM members.

\begin{figure}[t]
	\centering
	\includegraphics[width=0.8375\linewidth]{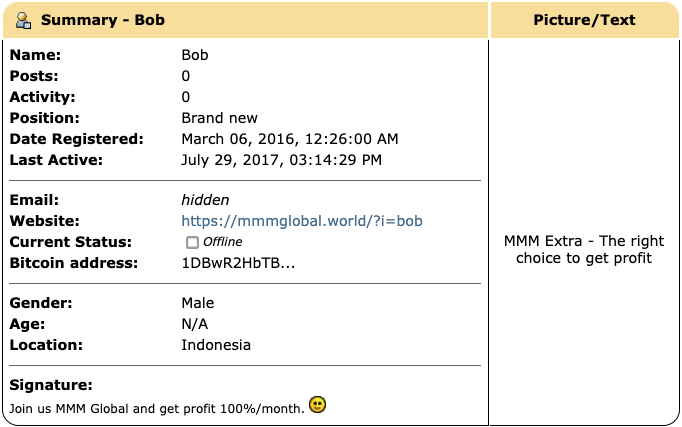}
	\caption{BitcoinTalk user profile showing a link to MMM.}
	\label{fig:bitcointalk_profile}
\end{figure}

We crawled 2.5M BitcoinTalk user profiles by downloading each profile page using a URL which is indexed by a user identifier. This resulted in 64 GB of profile pages in HTML format. We then parsed the pages to find Bitcoin addresses using regular expressions (\S\ref{sec:background_bitcoin}). This resulted in 39,321 addresses, each associated with a unique user profile. We identify MMM members based on the content of the website information declared in user profiles, as shown in Figure~\ref{fig:bitcointalk_profile}. In particular, if a user profile contains a hyperlink to a member profile on MMM, that is, a URL containing ``mmmglobal'' substring followed by a member identifier, we label the corresponding user as an MMM member. Out of 39,321 profile pages with Bitcoin addresses, we were able to label 24,200 profiles (61.54\%) as MMM members with 22,950 unique Bitcoin addresses. As some profiles declared the same Bitcoin address, 2,327 (9.61\%) in total, we decided to filter these out, ending up with 21,873 user profiles that are associated with unique MMM members and addresses. From now on, we use the terms ``user'' and ``member'' interchangeably but make the distinction when deemed necessary.

Out of 21,873 user profiles, a total of 3,861 users shared their age, 11,473 users shared their gender, and 2,588 users shared their location information on their profiles. We found that, on average, a user is 32.85 years old ($\sigma$=9.08). Gender distribution among users is skewed, with 75.18\% identifying as males. Finally, out of 80 different countries that users declared as their location, the top-3 are Indonesia (570 users), China (434 users), and then India (383 users).

Each profile page on BitcoinTalk includes timestamps of registration date and last activity. We found that 21,751 users (99.44\%) registered from Sep 2015 to Mar 2016. Only 285 users (1.30\%) have made one or more posts, where the most active user has made 377 posts. Also, 21,461 users (98.12\%) made their last activity between Sep 2015 and Mar 2016. On average, the activity period of a user is 8.56 days ($\sigma$=43.89), with the longest period of activity reaching 5.4 years. We manually inspected of a random sample of these user profiles and found that all of them were registered to promote MMM as part of MMM Extra scheme (see ~\S\ref{sec:background_mmm}), as shown in Figure~\ref{fig:bitcointalk_profile}.

Finally, we used WalletExplorer\footnote{\url{https://www.walletexplorer.com}} to identify the wallet to which a member address belongs, along with the wallet's label when available. We removed all addresses that belong to wallets which are labelled as non-exchange services, such as mining, gambling, and mixing. We do this filtering because such addresses are owned by services themselves, not members, and are not used for personal transactions. In total, 12,104 addresses belonged to unlabelled wallets, 3,632 addresses belonged to exchanges, that is, online wallets, and 6,137 addresses belonged to non-exchange services. As such, we ended up with 15,736 addresses, each associated with a unique MMM member. We refer to these addresses as ``Ponzi addresses.''

\subsection{Transactions}
\label{sec:dataset_transactions}

We used BlockSci~\cite{kalodner2017blocksci} v0.5.0 and Bitcoin Core v0.16.0 to collect all transactions which include any of the Ponzi addresses as inputs, outputs, or both. This resulted in 1,165,383 transactions. We then divided the transactions into three types depending on where Ponzi addresses appear in a transaction, as follows:
\begin{enumerate}
\item Deposits: Ponzi addresses appear only as outputs but never as inputs. Deposits move money into MMM.
\item Ponzi: Ponzi addresses appear as both inputs and outputs. Ponzi transactions circulate money within MMM.
\item Withdrawals: Ponzi addresses appear only as inputs but never as outputs. Withdrawals move money out of MMM.
\end{enumerate}

We removed deposits that represent mining rewards (i.e., Coinbase transactions), as they do not relate to MMM. We also removed all Ponzi transactions which involve only a single Ponzi address as an input and an output, as such transactions are likely payments to an unknown address, with the change sent back to the Ponzi address. We also removed other deposits and withdrawals that are not related to the overall Ponzi operation of MMM. In particular, we removed any deposit to a Ponzi address that is not followed by a Ponzi transaction which has the address as an input or an output. Similarly, we removed any withdrawal from a Ponzi address that is not preceded by a Ponzi transaction which includes the address as an input or an output. As such, we ended up with 291,172 deposits, 17,794 Ponzi transactions, and 113,987 withdrawals, adding up to a total of 422,953 transactions.

\subsection{External validation}
\label{sec:dataset_exnternal_validation}

We collected another dataset from Blockchain.com,\footnote{\url{https://www.blockchain.com}} a wallet and blockchain explorer service. Our goal is to further validate the association of the collected addresses to MMM by showing transactions linking addresses from both datasets. To achieve this, we repeated the collection procedures described above for the second source (Appendix~\ref{apndx:dataset}), and eventually found 187 transactions linking Ponzi addresses from BitcoinTalk and Blockchain.com. In particular, 139 transactions were sent from addresses in the BitcoinTalk dataset to addresses in the Blockchain.com dataset, with a total output value of 56,714 dollars. Moreover, 38 transactions were sent in the other direction, with a total output value of 47,145 dollars.

\section{Analysis}
\label{sec:analysis}

Next, we present a method with a set of metrics to quantitatively analyze Ponzi schemes on Bitcoin from four aspects, namely their lifecycle, Ponzi operation, externalities, and geopolitics. At the same time, we apply this method to MMM and characterize its operation for the first time. We also summarize the metrics in Appendix~\ref{apndx:metrics}.

\subsection{Lifecycle}
\label{sec:analysis_lifeycle}

Overall, MMM was operational for 2,111 days, or $\approx$5 years and 9.5 months, where the first transaction occurred on Oct 1, 2013, and the last transaction occurred on Jul 14, 2019. However, 98.58\% of all transactions occurred in a 3-year period, starting on Sep 1, 2014, and ending on Aug 31, 2017. We therefore restrict our analysis to this period and focus on measuring the volume of transactions and the corresponding flow of money.

\subsubsection{Volume.}
\label{sec:analysis_lifecycle_volume}

We analyze the lifecycle of a scheme in terms of its daily transaction volume (DTV), which is the number of deposit, Ponzi, and withdrawal transactions on any given day. In particular, we use DTV as a metric to model the lifecycle in the following three consecutive phases, which are formally defined in Appendix~\ref{apdx:formal_model}:
\begin{enumerate}
\item Bootstrap: The DTV grows steadily with a relatively steeper increase leading up to hyperoperation.
\item Hyperoperation: This phase consists of a number of cycles. In each cycle, the DTV grows nonlinearly, reaches a peak, and then decays nonlinearly.
\item Collapse: After that last decay of the last cycle in hyperoperation, the DTV decays steadily and approaches zero.
\end{enumerate}

\begin{figure}[t]
	\centering
	\includegraphics[width=0.75\linewidth]{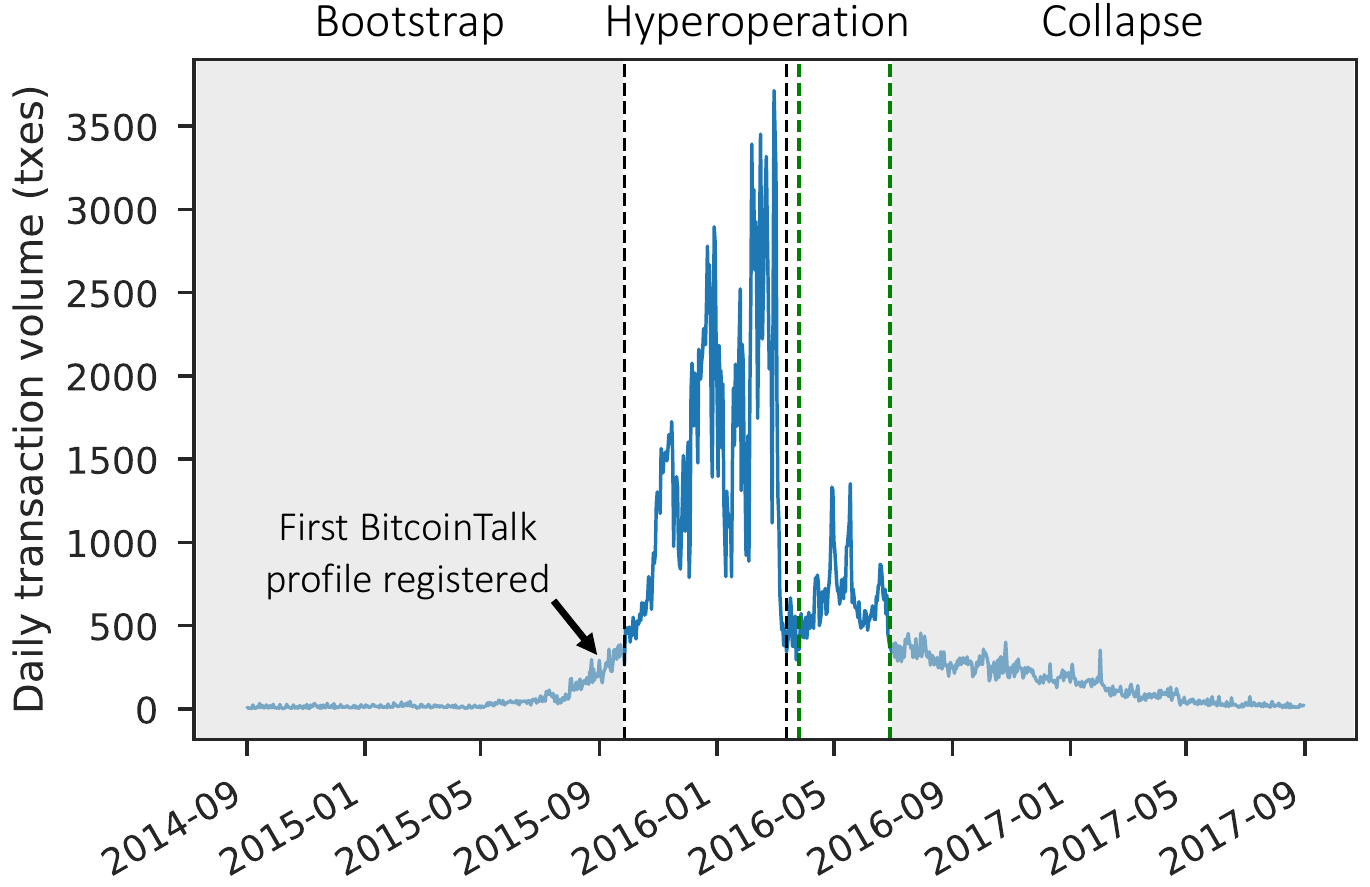}
	\caption{Daily transaction volume (DTV).}
	\label{fig:analysis_transaction_volume}
\end{figure}

We applied the definitions to MMM and ended up with the phases highlighted in Figure~\ref{fig:analysis_transaction_volume}. As shown in the figure, the hyperoperation phase started on Sep 27, 2015, a few days after the first MMM member registered her user profile on BitcoinTalk, and went through two cycles before ending on Jun 28, 2016. The first cycle peaked on Feb 29, 2016, with a DTV value of 3,714 transactions, after which it ended coinciding with reports announcing that MMM stopped paying its members in Apr 2016~\cite{collapse}. The second cycle, on the other hand, peaked on May 18, 2016, with a DTV value of 1,353 transactions, which is less than half its value in the previous cycle.

The objective behind modeling the lifecycle in phases is to analyze the scheme when the Ponzi operation is in full gear, which is captured by the hyperoperation phase, as presented in~\S\ref{sec:analysis_ponzi_operation}. While it is possible for a Ponzi scheme to resurrect and go through the three phases multiple times before stopping its operation completely, we did not observe this behavior with MMM.

\subsubsection{Money flow.}
\label{sec:analysis_moneyflow}

Given the lifecycle model defined above, we can measure how much money flows to and from the members of a Ponzi scheme in each of its phases. To accomplish that, we define a metric called the daily money flow (DMF-$f$) as the $f$-aggregate of input or output values of Ponzi addresses which appear in the transactions of a given day, where $f$ is an aggregate function like the sum, the mean, or the max. The values to consider for aggregation depend on the transaction type in which a Ponzi address is an input or an output, as defined in \S\ref{sec:dataset_transactions}.

\begin{figure}[t]
    \centering
	\includegraphics[width=0.75\linewidth]{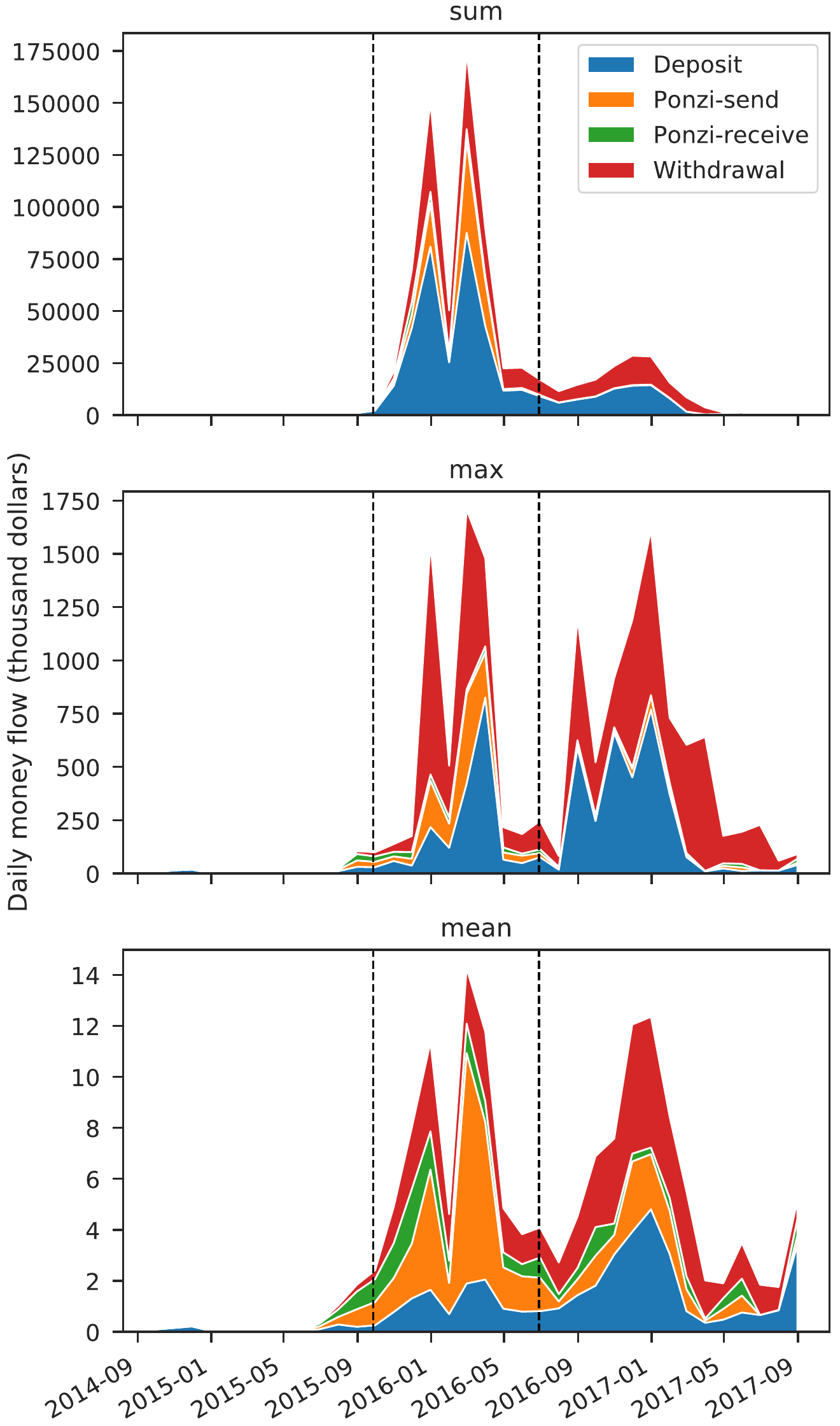}
	\caption{Daily money flow (DMF).}
	\label{fig:money_flow}
\end{figure}

\begin{figure}[t]
    \centering
	\includegraphics[width=0.75\linewidth]{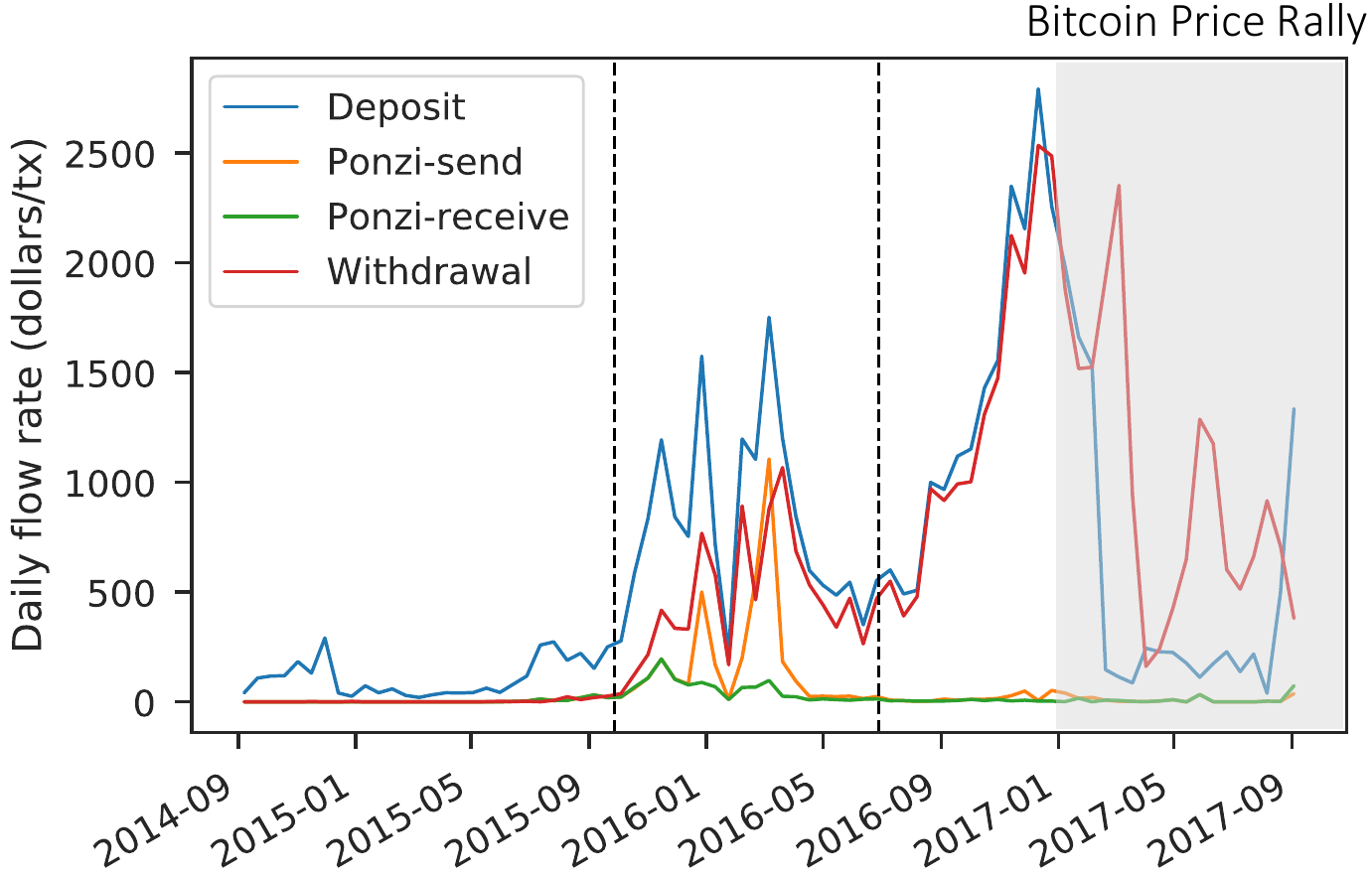}
	\caption{Daily flow rate (DFR).}
	\label{fig:dor}
\end{figure}

As shown in Figure~\ref{fig:money_flow}, most of the money flow in MMM occurred in the hyperoperation phase, with a peak DMF-sum value exceeding 150M dollars.\footnote{BlockSci uses CoinDesk to retrieve the price of Bitcoin in dollars on any given day.} The bootstrap phase, on the other hand, had an insignificant flow, with a peak DMF-sum value barely reaching 500K dollars. Unlike earlier phases, the collapse phase is dominated by flows that are attributed to a relatively small number of deposits and withdrawals with exceptionally large values, as evident by the difference between DMF-max and DMF-mean. This also indicates that MMM members have stopped participating in the scheme, with nearly no Ponzi-related money flow.

To measure the average contribution of a transaction to the daily money flow, we define a metric called the daily flow rate (DFR) as the DMF-sum divided by the DTV, which is the average dollar amount of a transaction per day. Figure \ref{fig:dor} shows the DFR in MMM, with a rapid increase in the collapse phase leading up to early 2017. We think this could be attributed to the 2017 Bitcoin price rally~\cite{btc_rally}, or the reuse of the Ponzi addresses in other, possibly illicit, activities.

\subsection{Ponzi Operation}
\label{sec:analysis_ponzi_operation}

We now focus on the hyperoperation phase and analyze Ponzi transactions in terms of income inequality, return on investment, and member classification. In MMM, a total of 8,680 Ponzi addresses appeared in Ponzi transactions during this phase.

\subsubsection{Income Inequality.}

In a Ponzi scheme, one expects a small number of members to receive most of the money from its daily operation. To measure the distribution of income across member, we use a metric called the Gini index~\cite{gastwirth1972estimation}, which is often used to measure income inequality in a population. A higher Gini index indicates greater inequality, with high income individuals receiving much larger percentages of the total income of the population.

We calculate the daily Gini index (DGI) as follows. First, on each day, we calculate the sum of output values for each Ponzi address in the corresponding Ponzi-receive transactions. We call this sum the daily net income (DNI) of an address. We then calculate the DGI as half of the relative mean absolute difference of DNI values across all Ponzi addresses. The value of DGI is between 0 and 1, where 0 indicates perfect equality in which every member has the same income, and 1 indicates perfect inequality in which only one member receives all the income.

\begin{figure}
    \centering
	\includegraphics[width=0.75\linewidth]{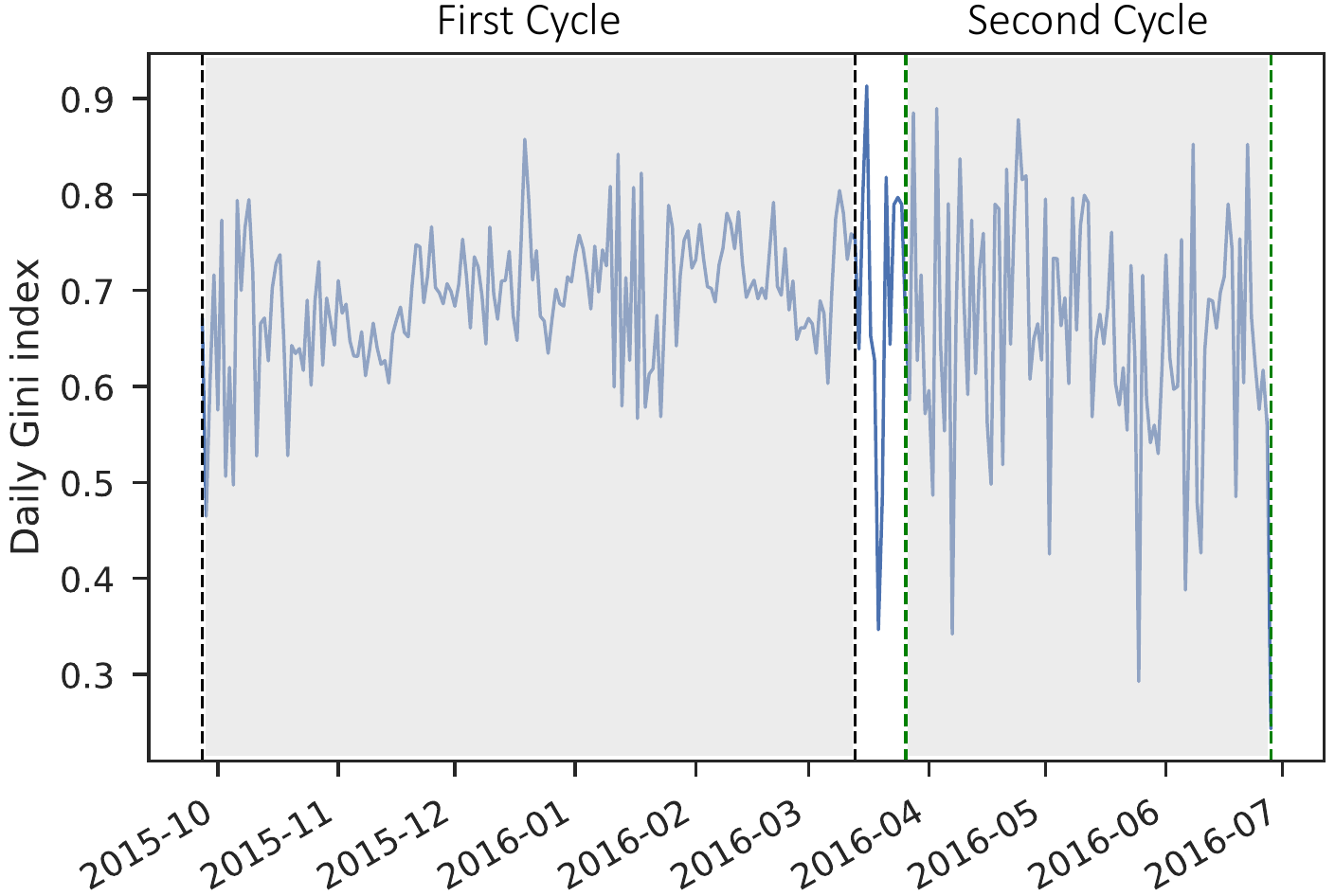}
	\caption{Daily Gini index (DGI).}
	\label{fig:daily_gini}
\end{figure}

As shown in Figure \ref{fig:daily_gini}, the DGI of MMM has an average of 0.68 and is nearly always higher than 0.35. As a reference, the Gini index of the UK in the financial year ending 2019 is 32.5\%~\cite{uk_gini}. This makes it one of the world's most unequal rich countries where the richest 10\% receive more than 40\% of the total income. As such, the DGI captures one aspect of MMM's Ponzi operation, in which a high concentration of income is received by a small number of members.

\subsubsection{Return on Investment.}

A Ponzi scheme generates returns on investments (ROIs) for older members by acquiring investments from new members, ultimately leading to a scenario where one member's profit is another member's loss. As such, a Ponzi scheme resembles a zero-sum game, where the sum of profits and losses among all members on any given day is equal to zero. To measure this effect, we first define a new metric for an address, called the daily net worth (DNW), as the difference between the address's DNI and the daily net spending (DNS), where DNS is the sum of its input values in each Ponzi-send transaction on any given day. A negative DNW value means the member associated with the address has lost money, while a positive DNW value means the member has made money. Next, we define a new metric called the daily net difference (DND) as the sum-aggregate of DNW values across all addresses. In concept, the DND value should be equal to zero and a non-zero value indicates that some money has been sent to or received from unknown addresses in Ponzi transactions. While these unknown addresses could be associated with the Ponzi scheme, we cannot simply include them without verification from other sources.

\begin{figure}
	\centering
	\includegraphics[width=0.75\linewidth]{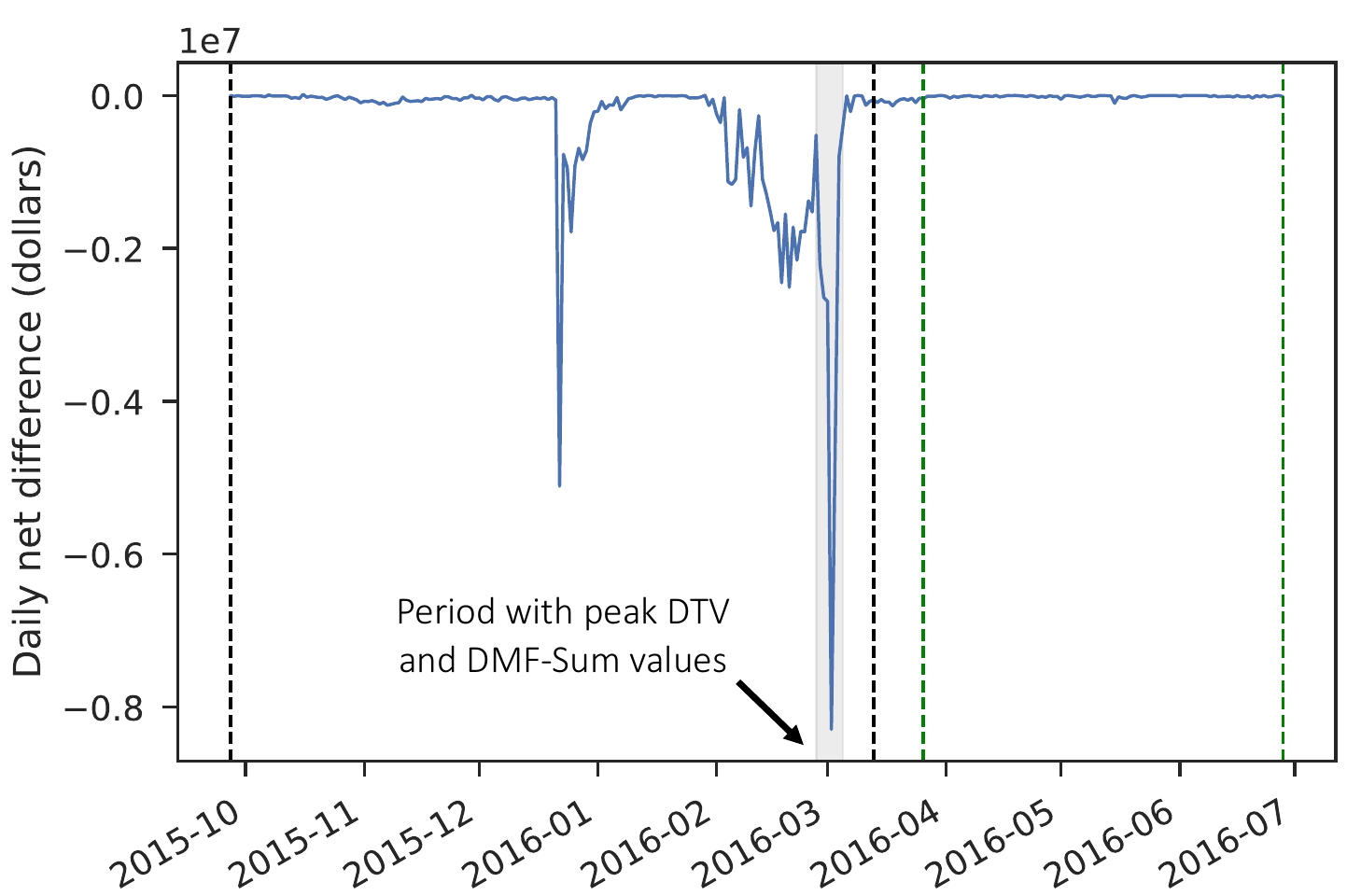}
	\caption{Daily net difference (DND).}
	\label{fig:daily_sum}
\end{figure}

As shown in Figure~\ref{fig:daily_sum}, the DND of MMM indicates that it exhibits a Ponzi zero-sum ROI model for most of its operation. However, there are occasions when money has been sent from Ponzi addresses to unknown ones, effectively ``leaking'' out from MMM. To measure the impact of this leakage, we calculate the total DND, which is the sum of DND values in the hyperoperation phase. If the total DND value is positive, then money is ``pumped'' into the Ponzi scheme. Otherwise, money is leaked out from the scheme. We also calculate the total absolute DND, which is the sum of absolute DND values in the hyperoperation phase. For MMM, we found a total DND value of -73.8M dollars and an absolute DND value of 73.9M dollars, which means most of the money is leaked, not pumped, and that the metrics we measured, those that involve Ponzi transactions in particular, represent lower-bound estimates of their actual values.

The lowest DND value, which is dated on Mar 2, 2016, is about eight million dollars and coincided with both the highest DTV value, shown in Figure~\ref{fig:analysis_transaction_volume}, and the highest DMF-sum value of the first cycle in the hyperoperation phase, shown in Figure~\ref{fig:money_flow}. This is also one example where a metric, such as DMF, excludes a significant sum of money due to unknown addresses in Ponzi transactions. The lowest DND value corresponds to 346 Ponzi transactions that include 9,167 unknown addresses with no intersection with those from the first peak, where only 0.19\% of the total input value and 0.01\% of the total output value were associated with Ponzi addresses. The second lowest DND value, on the other hand, corresponds to 178 Ponzi transactions that include 3,780 unknown addresses, where only 0.2\% of the total input value and 0.01\% of the total output value were associated with Ponzi addresses.

\subsubsection{Member Classification.}
\label{sec:scam}

While income inequality, as measured by the DGI, and a zero-sum ROI, as measured by the DND, indicate a Ponzi operation, these metrics do not answer the question: who are the victims and their likely scammers? We next address this question by applying profitability analysis to Ponzi transactions.

First, we define a new metric for an address on day $d_i$ of hyperoperation, called the total net worth (TNW($d_i$)), as the sum of its DNW values from the first day of hyperoperation, $d_1$, until day $d_i\le d_n$, where $d_n$ is the last day of hyperoperation. After that, we normalize TNW($d_i$) by dividing its value by the number of Ponzi transactions which involve the address as an input or an output between $d_1$ and $d_i$. We refer to this normalized metric by TNW-norm($d_i$).

On every day $d_i$, we rank all Ponzi addresses of the scheme based on their TNW-norm($d_i$) value in a descending order. As such, we can classify each address under the following rules:
\begin{enumerate}
\item TNW-norm($d_i$)$<$0: Associated member is a victim, as the member has lost money.
\item TNW-norm($d_i$)$\ge$0: Associated member is not a victim, as the member has not lost money or has made some money.
\item TNW-norm($d_i$)$\gg$0: Associated member is likely a scammer, as the member has made disproportionately more profit.
\end{enumerate}

\begin{figure}
	\centering
	\includegraphics[width=0.75\linewidth]{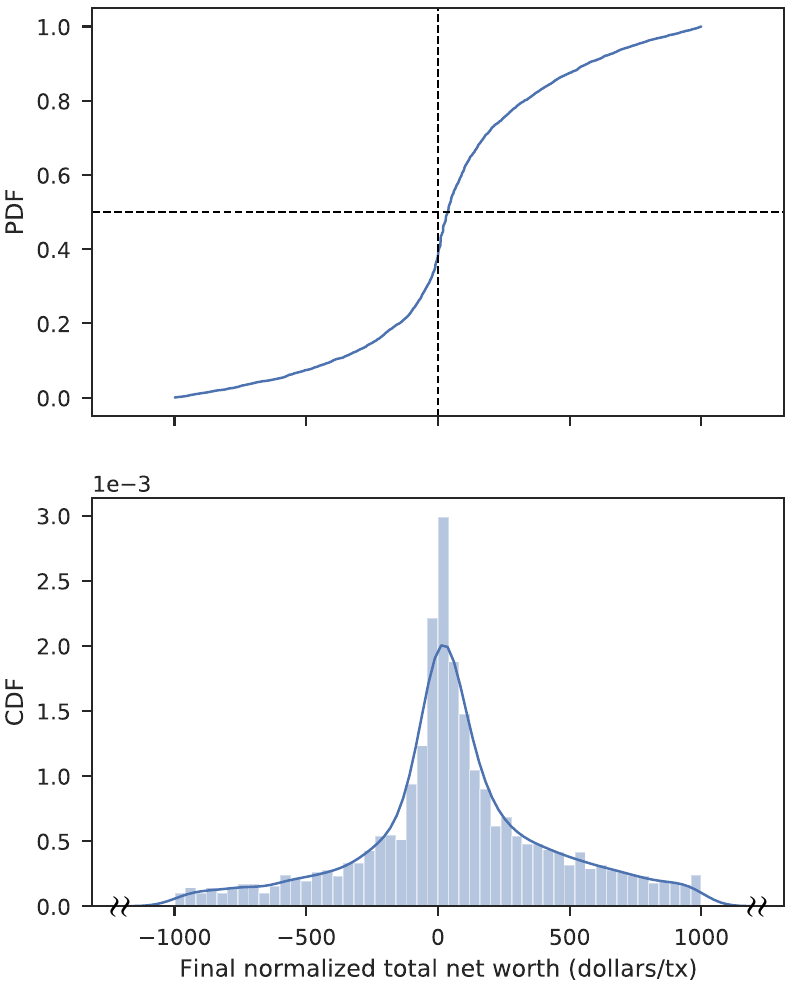}
	\caption{Final normalized total net worth (TNW-norm($d_n$)).}
	\label{fig:normalized}
\end{figure}

To get an end-result view of member classification, we calculate TNW-norm($d_n$), which is the final TNW-norm value for each Ponzi address in the scheme at the end of its hyperoperation phase. As shown in Figures~\ref{fig:normalized}, nearly 41\% of addresses have a value that is less than zero, which classifies their corresponding members as victims. This view, however, does not show how this percentage changes over time. Instead, we define a new metric for the scheme on day $d_i$ of its hyperoperation, called the cumulative victim ratio (CVR($d_i$)), as the ratio of addresses which are classified as victims on day $d_i$ according to rules (1) and (2) above.

\begin{figure}
	\centering
	\includegraphics[width=0.75\linewidth]{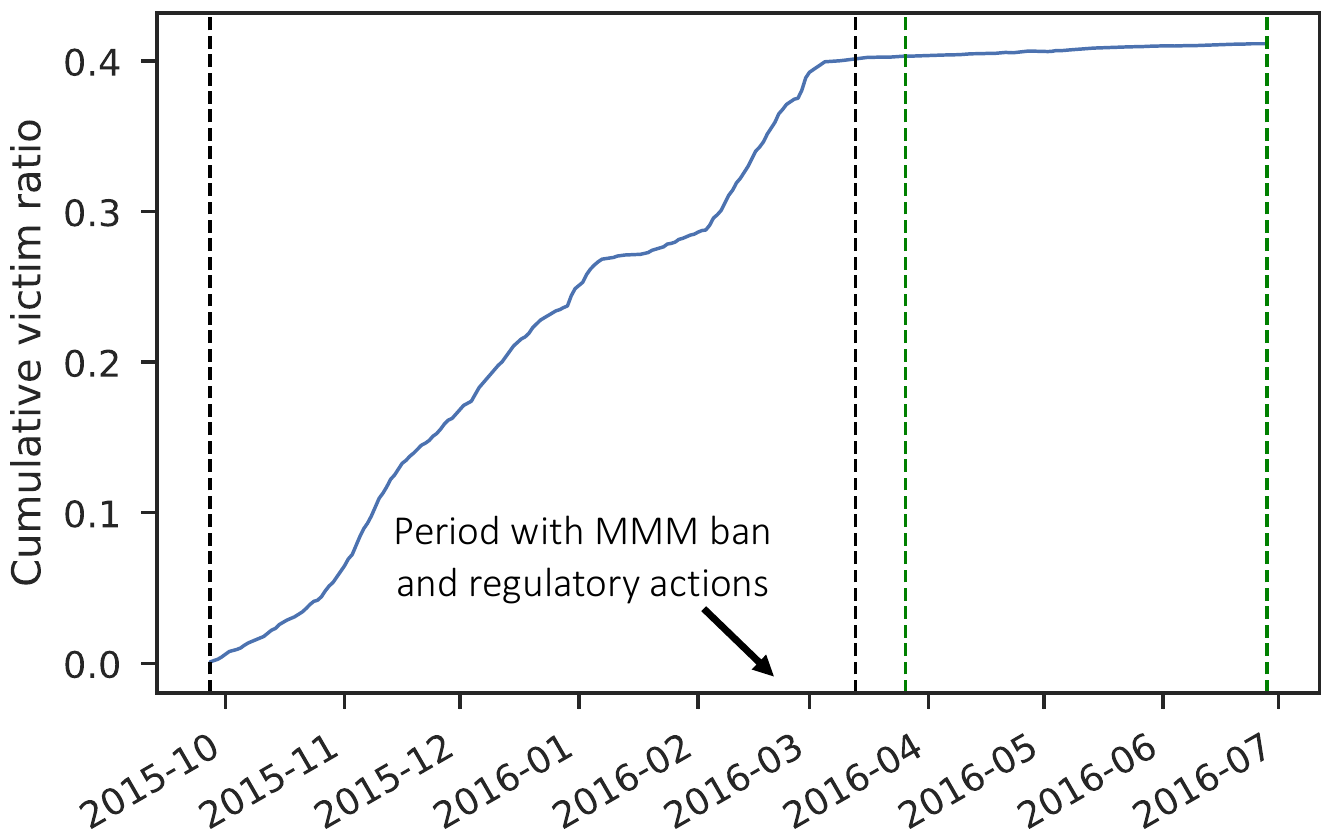}
	\caption{Cumulative victim ratio (CVR($d_i$)).}
	\label{fig:victims}
\end{figure}

As shown in Figure~\ref{fig:victims}, the CVR($d_i$) in MMM increases every day over a period of five months until it reaches 0.41, after which it hits a plateau in Mar 2016. This also coincides with the end of the first cycle of the hyperoperation phase, after which the scheme's DTV never fully recovers, leading up to a significantly lower peak in the second cycle, as shown in Figure~\ref{fig:analysis_transaction_volume}. The decrease in DTV also  maps to a reduced DMF-sum in the second cycle, especially from Ponzi transactions, as shown in Figure~\ref{fig:money_flow}. Together, the high CVR($d_i$) value and the low DTV and DMF-Sum values, indicate that victims have realized they are being scammed and have therefore reduced their participation in MMM. Interestingly, this also coincides with the ban of MMM in China~\cite{china} and the regulatory actions taken in South Africa against MMM in late Feb 2016~\cite{sa}.

To identify likely scammers, we used the following procedure: First, on each day $d_i$ during the hyperoperation phase, we find a list of top-10 Ponzi addresses which have the highest TNW-norm($d_i$) values of the day. After that, for each month, we select the Ponzi addresses that made it to any of the daily top-10 lists of the month. Finally, we classify a Ponzi address as a likely scammer address if it consistently appears in each month, as the member associated with such an address has gained significant profits, month after month, but never a loss. Table~\ref{tbl:scammers} presents the top-10 likely scammer addresses of MMM, ranked by their non-normalized, final TNW value. All but one of these addresses have a wallet size of 1, which means their corresponding members have used the same Ponzi address to perform one or more transactions during hyperoperation. Moreover, the user profiles associated with these addresses were registered from Oct 2015 to Feb 2016, which is just before the DTV peaked in the first cycle. The top deposits and withdrawals in terms of dollar value which are associated with some of these addresses, such the 1st and the 6th, involve common but unknown addresses, suggesting they might be connected. We study similar externalities in the upcoming section.

\begin{table*}[h!]
\caption{Top-10 likely scammer addresses of MMM.}
{\small
\begin{tabular}{l c c c l l l l}\toprule
 & Final TNW & Wallet size & Registration date & \multicolumn{2}{c}{Top deposit from} & \multicolumn{2}{c}{Top withdrawal to}\\
\cmidrule(lr){5-6}
\cmidrule(lr){7-8}
Address & (dollars) & (\# addresses) & (yyyy-mm-dd) & Address & Wallet & Address & Wallet \\ \midrule
1NGpnazTZR\dots & 764,988 & 1 & 2015-10-18 & {\em 1LVXa7xrFn\dots} & {\em 9877f5d228\dots} & {\em 1LVXa7xrFn\dots} & {\em 9877f5d228\dots} \\
1HM6Eo2GH2\dots & 137,490 & 1 & 2016-02-25 & 15cqdF756h\dots & d9e0f29d92\dots & 1FXHTFWdUXa\dots & de36bf7fec\dots \\
17NZuQpVnk\dots & 87,726 & 17 & 2015-11-14 & 1GvAtUaDA4\dots & 0747e64bec\dots & 1PLozjRDdof\dots & 00007d59f2\dots \\
1DpHywDhof\dots & 76,152 & 1 & 2015-11-22 & 1MBsJCWVkh\dots & 7066e604b0\dots & 1HWeepQYBE\dots & Huobi\\
1CXNqE5tRF\dots & 45,331 & 1 & 2015-12-10 & 17F2jPnJGT\dots & b09dc54caf\dots & 1HaitFo3qg\dots & df8997c4de\dots \\
176tD4SZRp\dots & 41,044 & 1 & 2015-11-19 & {\em 1LVXa7xrFn\dots} & {\em 9877f5d228\dots} & {\em 1LVXa7xrFn\dots} & {\em 9877f5d228\dots} \\
1Hr7fCEgEx\dots & 40,519 & 1 & 2016-02-25 & 1FTNFUPTkY\dots & ce8baa8e76\dots & 1GuPNdy5Db\dots & cc368bd729\dots \\
15sJ8Bta2U\dots & 38,297 & 1 & 2015-11-01 & 1KmNg57zJ\dots & Luno & 1KmNg57zJ\dots & Luno \\
1Nya6D7tUL\dots & 34,677 & 1 & 2016-02-13 & 1AZnsXDnF2\dots & 00027af4ad\dots & 1E2ajjqSMp\dots & 05eb115abd\dots \\
1DdLoYsYbC\dots & 20,259 & 1 & 2016-02-25 & 1KiA9UiNmu\dots & 31fd492691\dots & 1HDTpBqrv4\dots & 070655f16c\dots \\
\bottomrule
\end{tabular}
}
\label{tbl:scammers}
\end{table*}

\subsection{Externalities}
\label{sec:externalities}

Scheme externalities, such as connections with other cryptocurrency activities, are important for understanding how a scheme operates within the whole cryptocurrency ecosystem. We now focus on investigating where deposits come from and withdrawals go to by analyzing the wallets associated with these transactions. First, for each non-Ponzi address that appears as an input of a deposit or an output of a withdrawal, we get its wallet from WalletExplorer along with the wallet's label when available. We then group the wallets into eight categories, where each category specifies a service type to which the wallet belongs based on its labels, as follows:

\begin{enumerate}
\item Unknown: A personal wallet or unknown service.
\item Exchanges: A cryptocurrency exchange service.
\item Pools: A cryptocurrency mining pool.
\item Generic: A service that deals with Bitcoin such as payment processors, wallet services, and marketplaces.
\item Mixers: A cryptocurrency tumbling service. 
\item Gambling: A cryptocurrency casino or a gambling service.
\item Ponzi: A service promoting fraudulent schemes that promise high returns, including fake initial coin offerings (ICOs).
\item Darkweb: A Tor hidden service operating on the dark web.
\end{enumerate}
Finally, we calculate the percentage of deposits and withdrawals that fall under each one of these categories.

\begin{figure}
	\centering
	\includegraphics[width=0.9\linewidth]{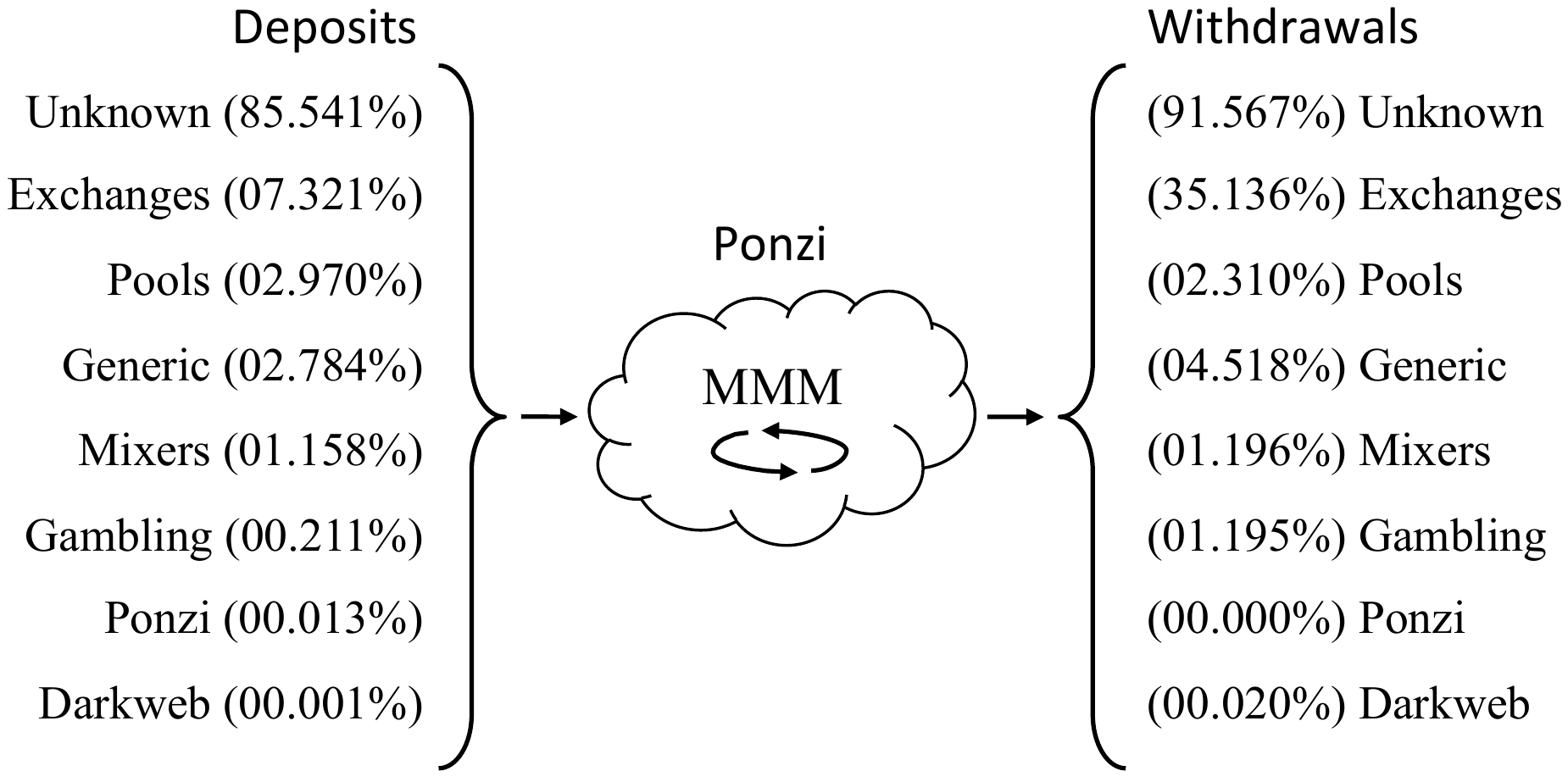}
	\caption{External cryptocurrency services and MMM.}
	\label{fig:service_categorization}
\end{figure}

As shown in Figure \ref{fig:service_categorization}, most of the deposits (85.54\%) and withdrawals (91.57\%) are associated with addresses that have unlabelled wallets, which means they might belong to personal, possibly member, wallets or simply to an unknown service. For those which we were able to identify, however, the highest percentages were associated with exchange services, with 4.8 times more withdrawals associated with exchanges than deposits. Moreover, there is a clear link between MMM and gambling services, which means they might be operated by the same entities. It is difficult to analyze why such links exist and for what reason, but one possible explanation is that certain MMM members might have tried to launder their profits using these services. The use of mixers or darkweb services further suggests that some MMM members are actively trying to conceal their identity, or have been involved in other illicit activities.

It is important to highlight that such an analysis is useful for law enforcement, as an investigation agency can work with regulated exchanges, which have to comply with know-your-customer (KYC) and anti-money laundering (AML) laws, to reveal the identities of likely scammers, once they cash out their profits.

\subsection{Geopolitics}
\label{sec:geopolitics}

Ponzi schemes target various countries and are particularly successful in developing or low-income economies~\cite{malkiel1999global, purba2017economics,
jack2018ponzi, sommers2010ponzi}. We now analyze the geopolitical reach of MMM and focus on how money flows between member countries. First, we use the location information shared by MMM members on their BitcoinTalk user profiles (\S\ref{sec:dataset_addresses}) to determine the country in which a member is located. We then group all Ponzi address that appear in Ponzi transactions by the country of their associated members. After that, for each group of addresses, we model it as a node in a directed graph called the geopolitical network. If any of the members who are located in a country have sent money to members who are located in another, we add an edge between the corresponding two nodes in the graph. The size of a node is relative to the number of members who are located in the corresponding country, and the thickness of an edge is relative to the amount of money which has been sent between the corresponding two countries, as per the direction of the edge.

\begin{figure}
	\centering
	\includegraphics[width=0.95\linewidth]{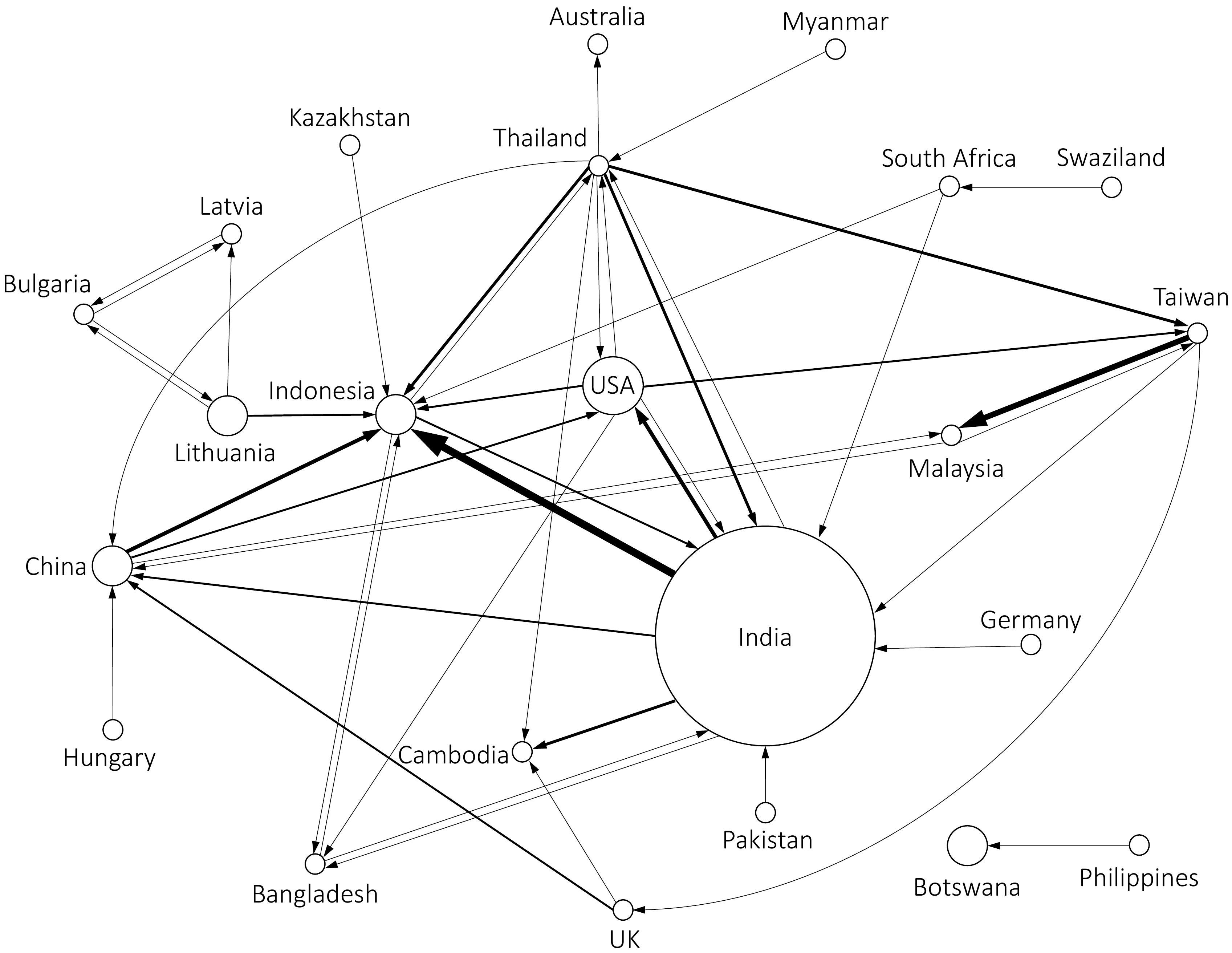}
	\caption{The geopolitical network of MMM.}
	\label{fig:country_graph}
\end{figure}

Figure \ref{fig:country_graph} shows the geopolitical network of MMM as a graph consisting of 23 nodes and 46 edges. It is clear that the scheme has extended its reach across the globe, from Africa, to Asia, Europe, and even Australia. MMM is also popular in the Caribbean and Brazil, which means it covers all the five continents. India and Indonesia are the two most prominent countries in MMM in terms of how much money they have circulated with others, making them central nodes in the graph. Countries like the USA, China, and Thailand are well-connected to each other, forming a small community in the graph along with India and Indonesia.

In terms of pairwise money flow, we found that the largest flow in MMM was between India and Indonesia, which was almost ten times the amount between Thailand and Taiwan, the pair with the smallest money flow. Moreover, it turns out that this pairwise money flow is highly asymmetrical. For example, members in Indonesia have received 12 times more money than they have sent to those in India. We also found that Indonesia, the highest receiving country, had received four times more money than the second-highest receiving country, Malaysia. These results suggest a few hypotheses, one of which is that Indonesia might be a central hub for MMM operators. This observation is supported by the fact that a significant amount of money is withdrawn from MMM through an Indonesian Bitcoin exchange (\S\ref{sec:externalities}).

\section{Discussion}
\label{sec:discussion}

This work, while having a number of limitations, has several implications for issues related to cryptocurrency regulations, economy, and user privacy. Next, we discuss these issues in detail.

\subsection{Implications}

First of all, it is important for governments to protect their people against financial e-crimes, such as operating a Bitcoin Ponzi scheme. During the first two quarters of 2019 alone, more than 4.26 billion dollars were lost due thefts, scams, and other misappropriation of funds from cryptocurrency users and exchanges~\cite{ciphertrace2019cryptocurrency}. The complicated nature of Ponzi schemes and the sophistication of Bitcoin's pseudo-anonymous privacy model have created a thriving environment for criminals to engage with various illicit activities. Moreover, the lack of effective mechanisms to identify cryptocurrency fraud sparked an urgent need for governments and regulators to come up with solutions. Fraud protection is part of the mandate of many financial regulators, and we believe that the analytical method and the set of metrics presented herein could be useful for customer protection and regulatory compliance. Combined with data mining and machine learning techniques (\S\ref{sec:related}), our investigation shows that new tools can be developed for better feature extraction, risk scoring, and automatic cryptocurrency fraud detection~\cite{boshmaf2019blocktag}. Such tools can be used by financial e-crime investigators and cryptocurrency exchanges to block transactions involving illicit activities.

Second, cryptocurrency fraud can cause significant damage to local economies. As we showed in \S\ref{sec:geopolitics}, the impact on the economy of several countries is a matter of concern. This could become a serious threat to monetary sovereignty, especially for developing countries, which are often targeted by Ponzi schemes. Therefore, it is important for governments to continuously monitor and prevent illicit cryptocurrency activities to protect their people and economy, and to thwart other issues, such as money laundering.

Third and last, the feasibility of this study raises a user privacy issue. While the identification of scammers is important, we showed that nearly half of MMM members are victims (\S\ref{sec:analysis_ponzi_operation}). These innocent people might not be privacy-savvy and are not aware of the risks associated with the participation in such activities. At a certain point, unveiling a Ponzi scheme requires deanonymization of user identities, which could expose the innocent ones to privacy breaches. For example, these users might have their names associated with financial fraud which could affect their credit history negatively. Moreover, sophisticated deanonymization techniques could extend to other Bitcoin addresses in their wallets, exposing more information about their financial activities~\cite{jawaheri2018small}. 

\subsection{Limitations}
\label{sec:limitations}

In what follows, we discuss the main limitations of the study and the steps we took to reduce their impact on the results when possible.

\subsubsection{Data collection}  We were able to collect a significant amount of Bitcoin addresses that are associated with MMM members mainly due to its aggressive marketing strategy. However, these addresses and their corresponding online identities are scraped from public user profiles that are not 100\% reliable. In a system where users are hidden behind pseudonyms, it is difficult to establish an absolute, complete ground truth. Nevertheless, we followed a strict validation procedure in order to filter out noise and irrelevant addresses or transactions, as described in \S\ref{sec:dataset}.

The zero-sum analysis presented in \S\ref{sec:analysis_ponzi_operation} also serves as a validity test for our dataset. In particular, it helps in quantifying the scale and impact of missed or excluded Ponzi addresses and their corresponding transactions. In particular, Figure \ref{fig:daily_sum} shows that for the significant majority of the hyperoperation phase, the Ponzi addresses included in our dataset yield Ponzi transactions with a near-zero-sum ROI, which denotes a relatively complete coverage of MMM's Ponzi operation among the considered members. At two significant instances, however, we acknowledge that the collected Ponzi addresses, and hence corresponding Ponzi transactions, were not complete and resulted in money leaks. Given that DNW-Sum in Figure~\ref{fig:daily_sum} is almost aligned with zero most of the time, it is safe to assume that the impact of the missing addresses and transactions is minimal. We would like to note that it is also possible that entities who sent money to unknown addresses might have deliberately done so for various reasons, such as paying someone or a service in the same Ponzi transaction, even though those unknown addresses do not belong to MMM.

To this end, our analysis shows that the collected dataset, while likely incomplete in terms of considering every MMM member and transaction, is still representative of MMM's operations, as we were able to capture its underlying Ponzi nature. One could increase the size of the dataset by crawling more public sources (Appendix~\ref{apndx:dataset}) or including more addresses using inference heuristics~\cite{meiklejohn2013fistful}. Such heuristics include adding new addresses by transitivity, where an unknown address receives money from and sends money to Ponzi addresses, or by closure, where the unknown address is grouped under a wallet to which a Ponzi address belongs. We do not use these heuristics simply because they bias and inflate the results~\cite{jawaheri2018small,ermilov2017automatic}.

\subsubsection{Analysis} From an analysis standpoint, we think that the temporal times-series analysis we performed herein does not provide a holistic view of a Ponzi scheme, leaving some interesting aspects undiscovered. We mainly studied events and money flow over time, but we did not study relationships among MMM transactions and members. Studying the scheme as a transaction or user graph could potentially reveal new characteristics that provide a broader understanding of underlying interactions. As such, including graph analysis could further enrich the investigation, and we leave it for a future work.

\section{Related Work}
\label{sec:related}

Financial fraud is a centuries old phenomenon that researchers have been trying to understand and resolve~\cite{fich2007financial, smith2010madoff, reuter2005chasing, reurink2016financial}. With the recent advent of cryptocurrencies, especially Bitcoin~\cite{nakamoto2008bitcoin, baek2015bitcoins, bohme2015bitcoin}, these currency networks have inadvertently become platforms for an array of fraudulent and unlawful schemes~\cite{tasca2015digital, nikolic2018finding, nabilou2018ignorance}. However, Bitcoin's main goal was and is still to decentralize and liberate traditional banking~\cite{marcel2017fintech, yermack2015bitcoin, grinberg2012bitcoin}. For people with privacy expectations, Bitcoin also provided pseudo-anonymity so that identities are masked to a significant extent~\cite{reid2013analysis, moser2013anonymity}.

Bitcoin's pseudo-anonymity has been used by criminals as a layer to hide behind and evade law enforcement and regulators~\cite{money_laundering, bohr2014uses}. Previous work has shown that services, such as mining scams, Ponzi schemes, scam wallets, and fraudulent exchanges are prevalent in the cryptocurrency ecosystem~\cite{vasek2015there, bohr2014uses}. For example, more than half of the bitcoin exchanges either got hacked, were forced to close, or have scammed users~\cite{moore2013beware}.  

Money laundering~\cite{reuter2005chasing} and Ponzi schemes~\cite{moore2012postmodern} have long existed before Bitcoin. Money laundering in Bitcoin, however, has received more attention from the research community~\cite{money_laundering, bitcoin_laundry_services}. Money laundering can be used for various reasons, such as evading taxes~\cite{quirk1997money} or hiding money accumulated using unlawful means~\cite{reuter2005chasing}. The main difference between money laundering and a Ponzi scheme is that the later targets ordinary people to steal their money~\cite{artzrouni2009mathematics}. To the best of our knowledge, we are the first to investigate a real-world Ponzi scheme on Bitcoin.

A recent, closely-related work proposed a data mining technique to detect Bitcoin addresses that are involved in Ponzi schemes~\cite{bartoletti2018data}. In their study, the authors provide an automatic analysis of Ponzi schemes over Bitcoin using supervised learning algorithms. In particular, they manually collected Bitcoin addresses 
from different forums and discussion boards. Then they used a clustering heuristic~\cite{meiklejohn2013fistful} in order to expand the collected dataset. They also define a set of features to be used in Bitcoin addresses classification. Furthermore, by using different classifiers, they tried to detect features from other addresses related to Ponzi schemes. Their best classifier was able to classify 31 out of 32 Ponzi schemes, with 1\% false positives. Interestingly, the authors have explicitly excluded MMM, as its transaction history was too complicated for automated detection using the simple features they used. In contrast, in our work, we do not aim to identify Ponzi schemes but rather analyze their operation using a set of metrics. We show, in the case of MMM, that more sophisticated features, which could be derived from the metrics, are needed to effectively capture the Ponzi operation.

\section{Conclusion}

We presented the first in-depth analysis of a real-world Ponzi scheme on Bitcoin. We analyzed 432K Bitcoin transaction involving 16K addresses associated with MMM. We showed that MMM exhibits many properties of a Ponzi scheme, including a short life span, income inequality, a zero-sum return on investment model, an increasing number of victims over time, involvement of exchanges, and a global reach.

\section*{Acknowledgements}
We would like to thank the people at the Cybersecurity Initiative for Blockchain Research (CIBR) for their help and feedback.\footnote{For latest research outcomes, please visit \url{https://qcri.github.io/cibr}}

\bibliographystyle{refs}
\bibliography{refs}

\appendix

\section{Supplementary Dataset}
\label{apndx:dataset}

We crawled Blockchain.com to search for public tags of Bitcoin addresses which are associated with MMM. After cleaning, we ended up with 181 addresses, each associated with a unique MMM tag. We then used BlockSci to extract the corresponding transactions. After cleaning, we ended up with 7,322 transactions.

\subsection{Addresses}
\label{apndx:dataset_addresses}

Blockchain.com is one of the most popular wallet and blockchain explorer services. As of Feb, 2019, the platform hosted 33M user wallets and was used to perform more than 200B transactions in 140 countries. On Blockchain.com, users can tag a Bitcoin address with a short label and a URL by either submitting a hyperlink as a reference or digitally sign a random message using the private key of the address being tagged. This information is used to verify the ownership of an addresses and its tag, and to annotate the corresponding transactions in the blockchain. Based on the platform's popularity, we used it as a second source to collect Bitcoin addresses that are associated with MMM members.

We downloaded 600 address tag pages, each containing tag information of at most 50 addresses, by retrieving a page using a unique URL indexed by a tag identifier. There are also two tag types, submitted links and signed messages, that are represented by two unique identifiers in the URL. This resulted in about 40MB of unparsed address tag pages in HTML format. We then parsed the pages to extract Bitcoin addresses and their tag information, namely their short labels, URLs, and whether a tag is verified. We ended up with 30,009 tagged Bitcoin addresses, out of which 29,643 addresses (98.78\%) had verified tags.

We identify MMM tags by looking at their tag label. In particular, if the short label of a tag contains ``mmm'' substring, we mark the corresponding address as a candidate. Accordingly, out of 29,643 verified tagged addresses, we found 202 candidates (0.68\%). After manual inspection, we excluded 21 addresses, as they are not likely related to MMM (e.g., a tag with a short label ``hmmmm''). This resulted in 181 addresses that have MMM tags, out of which two tags were verified by submitted hyperlinks and 179 tags were verified by digitally-signed messages. We only consider these validated, verified MMM tags. We consider the corresponding addresses to be associated with MMM, as their tag labels include the scheme's name in addition to relevant contextual information, such as whether the tag represents a certain country, a global support fund, or a member, as shown in Table~\ref{table:appendix_blockchain_info_top10}.

\begin{table}
\centering
\caption{Top-10 most frequent MMM tags.}
{\small
\begin{tabular}{lr} \toprule
    Label & Frequency\\ \midrule
    mmm universe.help & 46\\
    mmm global & 13\\
    bonus from mmm universe.help & 9\\
    mmm indonesia & 6\\
    mmm nusantara & 4\\
    mmm china & 2\\
    mmm india & 2\\
    mmm indonesia & 2\\
    mmm philippines & 2\\
    mmm russia & 2\\
\bottomrule
\end{tabular}
}
\label{table:appendix_blockchain_info_top10}
\end{table}

\subsection{Transactions}
\label{apndx:dataset_transactions}

\begin{table*}
\centering
\caption{Summary of metrics}
{\small
\begin{tabular}{lll} \toprule
	Acronym & Name & Brief definition\\ \midrule
	DTV & Daily transaction volume & Number of deposit, Ponzi, and withdrawal transactions on any given day\\
	DMF-$f$ & Daily money flow & $f$-aggregate of input or output values of Ponzi addresses which appear in the transactions of a given day\\
	DFR & Daily flow rate & DMF-sum divided by DTV\\
	DNI & Daily net income & Sum of output values for a Ponzi address in the corresponding Ponzi-receive transactions on any given day\\
	DGI & Daily Gini index & Half of the relative mean absolute difference of DNI values across all Ponzi addresses\\
	DNS &  Daily net spending & Sum of the input values for a Ponzi address in each Ponzi-send transaction on any given day\\
	DNW & Daily net worth & Difference between DNI and DNS of a Ponzi address on any given day\\
	DND & Daily net difference & Sum-aggregate of DNW values across all Ponzi addresses\\
	TNW($d_i$) &  Total net worth & Sum of a Ponzi address DNW values from the first day of hyperoperation until a given day $d_i$\\
	TNW-norm($d_i$) & Normalized TNW($d_i$) & TNW($d_i$) divided by number of Ponzi transactions involving the address as an input or an output until day $d_i$\\
	CVR($d_i$) & Cumulative victim ratio & Ratio of addresses which are classified as victims on day $d_i$\\
\bottomrule
\end{tabular}
}
\label{table:appendix_metrics}
\end{table*}

We used BlockSci to collect all transactions which include any of the MMM addresses as inputs, outputs, or both. This resulted in 9,293 transactions. We then grouped the transactions into three types, namely deposits, Ponzi, and withdrawals, and removed invalid transactions, as discussed in~\S\ref{sec:dataset_transactions}. We ended up with 6,853 deposits, 97 Ponzi transactions, and 372 withdrawals, adding up to a total of 7,322 transactions.

\section{Metrics}
\label{apndx:metrics}

As a quick reference, we refer the reader to Table~\ref{table:appendix_metrics}, which presents a summary of the metrics defined and used in this paper.

\section{Formal model}
\label{apdx:formal_model}

Let $\texttt{vols}=\{v_0,\dots,v_n\}$ be the daily transaction volume of a Ponzi scheme where $v_i \ge 0$ is the volume on the $i$-th day. Our goal is to partition $\texttt{vols}$ intro three disjoint sets, each representing one of the three phases, namely bootstrap, hyperoperation, and collapse. To achieve this, it is enough to find the set that corresponds to the middle phase, as the partitioning preserves the ordering in $\texttt{vols}$.

In Listing~\ref{listing:get_phase}, we define the hyperoperation phase computationally by finding its start day, end day, and cycles. This phase consists of one or more cycles, each having its own start and end days. The first cycle is the one that has the largest daily transaction volume, also referred to as the first peak. The start day of the phase is the start day of the earliest cycle. Similarly, the end day of the phase is the end day of the latest cycle.

\begin{lstlisting}[label={listing:get_phase},float,caption={\textmd{\texttt{get\_phase(vols)}}},captionpos=t]
start_day = 0
end_day = len(vols)-1
phase = {
    'starts': end_day,
    'ends': start_day,
    'cycles': []
}
# first cycle
phase['cycles'].append(
    get_cycle(vols, cycles, start_day, end_day)
)
# cycles before the first
while True:
    cycle = get_cycle(
        vols, cycles,
        start_day,
        phase['cycles'][-1]['starts']
    )
    if valid_cycle(cycle, phase['cycles']):
        phase['cycles'].append(cycle)
    else:
        phase['starts'] = phase['cycles'][-1]['starts']
        break
# cycles after the first
while True:
    cycle = get_cycle(
        vols, cycles,
        phase['cycles'][-1]['ends'],
        end_day
    )
    if valid_cycle(cycle, phase['cycles']):
        phase['cycles'].append(cycle)
    else:
        phase['ends'] = phase['cycles'][-1]['ends']
        break
return phase
\end{lstlisting}

As described in Listing~\ref{listing:get_cycle}, a cycle is defined by the start day that is before the day of its peak and on which the volume is at most $1/g$-th of the peak, where $g$ is a growth factor. In other words, the start day is the latest day before the day of the peak after which the volume grows by a factor of $g$. The end day of the cycle is defined similarly, albeit after the peak day. Listing~\ref{listing:get_peak} shows how to find the peak volume of a period of time as specified by two days. 

\begin{lstlisting}[label={listing:get_cycle},float,caption={\textmd{\texttt{get\_cycle(vols, cycles, start\_day, end\_day)}}},captionpos=t]
peak_vol, peak_day = get_peak(vols, start_day, end_day)
cycle = {
    'starts': peak_day,
    'ends': peak_day,
    'duration': 0,
    'peak_vol': peak_vol,
    'peak_day': peak_day,
    'factor': get_factor(cycles, peak_vol)
}
# days before the peak
for day in range(cycle['peak_day'], start_day, -1):
    if vols[day-1] <=  cycle['peak_vol']/cycle['factor']:
        cycle['starts'] = day-1
# days after the peak
for day in range(cycle['peak_day'], end_day, 1):
    if vol[day+1] <=  cycle['peak_vol']/cycle['factor']:
        cycle['ends'] = day+1
if len(cycles) > 1:
    cycle['starts'] = max(cycle['starts'], cycles[-1]['ends'])
cycle['duration'] = cycle['ends'] - cycle['starts']
return cycle
\end{lstlisting}

\begin{lstlisting}[label={listing:get_peak},float,caption={\textmd{\texttt{get\_peak(vols, start\_day, end\_day)}}},captionpos=t]
peak_vol = 0
peak_day = 0
for day in range(start_day, end_day):
    vol = vols[day]
    if vol > peak_vol:
        peak_vol = vol
        peak_day = day
return peak_vol, peak_day
\end{lstlisting}

We define the growth factor such that the first peak is an order of magnitude higher than the volume of the last day of the previous cycle or phase, if there is only one cycle, as described in Listing~\ref{listing:get_factor}. We chose a factor of 10 in order to model nonlinearity in the growth during the hyperoperation phase. For the second cycle in this phase, the growth factor is defined as the factor of the first cycle multiplied by the ratio between the two peaks. This makes sure that this cycle is defined over a period that is relative to the change in peaks. As such, this definition is applied recursively for proceedings cycles.

\begin{lstlisting}[label={listing:get_factor},float,caption={\textmd{\texttt{get\_factor(cycles, peak\_vol)}}},captionpos=t]
if len(cycles) == 0:
    return 10
else:
    last_cycle = cycles[-1]
    vol_ratio = peak_vol / last_cycle['peak_vol']
    return max(1, last_cycle['factor'] * vol_ratio)
\end{lstlisting}

Finally, we keep adding new cycles until we reach an invalid one, which is determined based on its duration, as described in Listing~\ref{listing:valid_cycle}.

\begin{lstlisting}[label={listing:valid_cycle},float,caption={\textmd{\texttt{valid\_cycle(cycle, cycles)}}},captionpos=t]
low = cycles[-1]['duration'] / cycle['factor']
valid_low = cycle['duration'] >= low
valid_high = cycle['duration'] < cycles[-1]['duration']
return valid_low and valid_high
\end{lstlisting}

\end{document}